# PSIRNet: Deep Learning–based Free-breathing Rapid Acquisition Late Enhancement Imaging

## Authors

Arda Atalik, MPhil[1, 2, 3, 4]; Hui Xue, PhD[1]; Rhodri H. Davies, MD, PhD[5, 6]; Thomas A. Treibel, MD[5, 6]; Daniel K. Sodickson, MD, PhD[2, 3]; Michael S. Hansen, PhD[1]; Peter Kellman, PhD[1]

## Affiliations

1. Health Futures, Microsoft Research, Redmond, WA, USA
2. Center for Data Science, New York University, New York, NY, USA
3. Center for Advanced Imaging Innovation and Research (CAI$^2$R), Department of Radiology, NYU Grossman School of Medicine, New York, NY, USA
4. Bernard and Irene Schwartz Center for Biomedical Imaging, Department of Radiology, NYU Grossman School of Medicine, New York, NY, USA
5. Institute of Cardiovascular Science, University College London, London, UK
6. Barts Heart Centre, Barts Health NHS Trust, London, UK

**Summary Statement:** Deep learning–based free-breathing late gadolinium enhancement imaging produces diagnostic-quality phase-sensitive inversion recovery images from a single acquisition, matching or exceeding reconstruction quality using 8 to 24 motion-corrected averages.

## Key Points

- Single-shot PSIRNet reduced acquisition time by 8- to 24-fold compared to motion-corrected averaged reconstructions; it achieved statistically superior expert-scored image quality for dark blood late gadolinium enhancement and equivalent quality for bright blood and wideband variants.
- The network was trained, validated, and externally tested on 800,653 slices from 55,917 patients across multiple institutions, field strengths, scanner models, and software versions, demonstrating robust generalizability across bright blood, dark blood, and wideband late gadolinium enhancement variants.
- Inference required approximately 100 msec per slice using 10 gigabytes of graphics memory, compared with more than 5 sec for the conventional motion-corrected averaging pipeline, enabling near real-time reconstruction.

# Abstract

**Purpose:**

To develop and evaluate a deep learning–based method for free-breathing phase-sensitive inversion recovery (PSIR) late gadolinium enhancement (LGE) cardiac MRI that produces diagnostic-quality images from a single acquisition over two heartbeats, eliminating the need for 8 to 24 motion-corrected (MOCO) signal averages.

**Materials and Methods:**

Raw data comprising 800,653 slices from 55,917 patients, acquired on 1.5T and 3T scanners across multiple sites from 2016 to 2024, were used in this retrospective study. Data were split by patient: 640,000 slices (42,822 patients) for training and the remainder for validation (4,952 patients) and testing (8,143 patients), without overlap. The training and testing data were from different institutions. PSIRNet, a physics-guided deep learning network with 845 million parameters, was trained end-to-end to reconstruct PSIR images with surface coil correction from a single interleaved inversion recovery/proton density acquisition over two heartbeats. Reconstruction quality was evaluated using structural similarity index measure (SSIM), peak signal-to-noise ratio (PSNR), and normalized root mean square error (NRMSE) against MOCO PSIR references. Two expert cardiologists performed an independent qualitative assessment, scoring image quality on a 5-point Likert scale across bright blood, dark blood, and wideband LGE variants. Paired superiority and equivalence (margin = 0.25 Likert points) were tested using exact Wilcoxon signed-rank tests at a significance level of 0.05 using R version 4.5.2.

**Results:**

Both readers rated single-average PSIRNet reconstructions superior to MOCO PSIR for dark blood LGE (conservative $P = .002$); for bright blood and wideband, one reader rated it superior and the other confirmed equivalence (all $P < .001$). Inference required approximately 100 msec per slice versus more than 5 sec for MOCO PSIR.

**Conclusion:**

PSIRNet produces diagnostic-quality free-breathing PSIR LGE images from a single acquisition, enabling 8- to 24-fold reduction in acquisition time.

# Introduction

Late gadolinium enhancement (LGE) cardiac MRI is the clinical reference standard for assessing myocardial viability, infarction, and scar in both ischemic heart disease and many types of cardiomyopathies (1). Phase-sensitive inversion recovery (PSIR) enhances LGE by preserving the sign of longitudinal magnetization, which makes image contrast less sensitive to the choice of inversion time and simplifies clinical interpretation (2). However, conventional LGE imaging is highly time-consuming, averaging 15 minutes per study and occupying up to a third of the total scan time depending on the protocol. Accelerating this acquisition could significantly reduce cardiac MRI slot times or enable higher-resolution imaging within the same timeframe.

To eliminate the need for repeated breath-holds, free-breathing single-shot PSIR with non-rigid motion-corrected (MOCO) averaging has become widely adopted, delivering image quality comparable to breath-held segmented acquisitions (3). The MOCO PSIR method registers all inversion recovery (IR) images together to remove respiratory motion. After dropping the half of the heartbeats with the most different respiratory phases, the remaining MOCO IR images are averaged. The same process is repeated for proton density (PD) images. The final PSIR image is computed from the averaged IR and PD images. However, this approach typically requires 8 or more signal averages to recover an adequate signal-to-noise ratio (SNR), with wideband and dark blood protocols demanding up to 24 averages (4). Each average requires acquiring both an IR and a PD image over two heartbeats. The resulting scan time limits volumetric coverage, constrains achievable slice thickness, and reduces the time available for other sequences in the examination. The computational burden is also substantial: the motion correction, registration, and averaging pipeline requires several seconds per slice, which compounds over the multiple slices needed for full heart coverage.

Several deep learning methods have been proposed to accelerate LGE imaging, but prior work has largely focused on breath-held, segmented acquisitions at modest acceleration factors (5–9). Additionally, existing studies have typically relied on small, single-center cohorts, limiting generalizability across scanner vendors, field strengths, anatomical views, and LGE variants such as dark blood and wideband protocols. A method that generalizes across these dimensions while operating on the single-shot, free-breathing acquisitions used in routine clinical practice remains an unmet need.

In this work, we propose PSIRNet, an end-to-end, physics-guided deep learning method for free-breathing PSIR LGE cardiac MRI. Using a single interleaved inversion recovery/proton density (IR/PD) acquisition, PSIRNet reconstructs a PSIR image with surface coil correction through a series of learned Landweber iterations, each incorporating adaptive data-consistency guidance and deep learning–based refinement. The network was trained, validated, and externally tested on a large multi-institutional dataset, encompassing bright blood, dark blood (10), and wideband (11) LGE variants. We hypothesized that PSIRNet would produce diagnostic-quality PSIR images from a single acquisition with image quality equivalent to or exceeding that of conventional MOCO PSIR, thereby enabling 8- to 24-fold acceleration in acquisition time. We evaluated this hypothesis through quantitative image quality metrics and an independent qualitative assessment by two expert cardiologists.

## Materials and Methods

### Dataset

This retrospective study utilized LGE raw *k*-space data from 800,653 slices (55,917 patients) of accelerated single-shot parallel imaging acquisitions from 1.5T and 3T scanners across multiple institutions, encompassing a range of scanner models and software versions. Data acquired between 2016 and 2024 from the National Institutes of Health Cardiac MRI Raw Data Repository, hosted by the Intramural Research Program of the National Heart, Lung, and Blood Institute, were curated with the required ethical and/or secondary audit use approvals or guidelines permitting the retrospective analysis of anonymized data without requiring written informed consent for secondary usage for the purpose of technical development, protocol optimization, and/or quality control. No exclusion criteria were applied. The data were fully anonymized and were split at the patient level to prevent leakage in network training.

**Table 1.** Typical Acquisition Parameters of Bright Blood, Dark Blood, and Wideband MOCO PSIR LGE Protocols.

| | Bright Blood | Dark Blood | Wideband |
|---|---|---|---|
| Preparation | Inversion Preparation | Inversion Preparation and T2 Preparation | Wideband Inversion Preparation |
| Readout (single shot) | SSFP $FA_{IR} = 50°$ $FA_{PD} = 8°$ | SSFP $FA_{IR} = 50°$ $FA_{PD} = 8°$ | FLASH $FA_{IR} = 10°$ $FA_{PD} = 5°$ |
| Typical FOV (mm$^2$) | 360×270 | 360×270 | 360×270 |
| Typical resolution (mm$^3$) | 1.4×1.9×8 | 1.4×1.9×8 | 1.4×1.9×8 |
| Matrix size | 256×144, PAT 2 or 3 | 256×144, PAT 2 or 3 | 256×144, PAT 2 or 3 |
| Averages | 8 (16 beats) | 12–16 (24–32 beats) | 12–24 (24–48 beats) |
| T2 prep TE (msec) | N/A | 10–40 | N/A |
| TE / TR (msec) | 1.2 / 2.8 | 1.2 / 2.8 | 1.25 / 3.06 |

*Note.*—FA = flip angle, FLASH = fast low-angle shot, FOV = field of view, IR = inversion recovery, LGE = late gadolinium enhancement, MOCO = motion corrected, N/A = not applicable, PAT = parallel acquisition technique acceleration factor, PD = proton density, PSIR = phase-sensitive inversion recovery, SSFP = steady-state free precession, TE = echo time, TR = repetition time.

That is, all slices from a given patient were assigned to a single partition: 640,000 slices (42,822 patients) for training, 80,014 slices (4,952 patients) for validation, and 80,639 slices (8,143 patients) for external testing. For each slice, 8 to 24 single-shot IR and PD images were acquired during free breathing in an interleaved order: an IR image was acquired during the first heartbeat, followed by a PD image in the second heartbeat. We ensured there was no patient overlap across splits, and the training and external

**Table 2.** Patient Demographics in Each Dataset Split for Bright Blood, Dark Blood, and Wideband LGE Variants.

|  |  | Bright Blood | Dark Blood | Wideband |
|---|---|---|---|---|
| Training | No. of Slices | 618,681 | 13,850 | 7,469 |
|  | No. of Patients | 42,459 | 2,463 | 614 |
|  | Sex | 10,346M/6,279F/29U | 661M/288F/13U | 199M/103F/10U |
|  | Age | 56.0 (24.0) | 59.0 (21.0) | 59.0 (20.3) |
|  | Weight | 80.0 (25.0) | 80.0 (22.0) | 80.0 (24.3) |
|  | Height | 170.0 (15.0) | 170.0 (14.6) | 171.0 (13.3) |
| Validation | No. of Slices | 70,794 | 4,253 | 4,967 |
|  | No. of Patients | 4,596 | 466 | 377 |
|  | Sex | 2,774M/1,821F/1U | 254M/211F/0U | 222M/144F/0U |
|  | Age | 59.0 (23.0) | 64.0 (23.0) | 60.5 (23.0) |
|  | Weight | 79.4 (25.4) | 78.9 (27.8) | 83.0 (28.0) |
|  | Height | 170.2 (15.0) | 170.2 (15.8) | 172.0 (13.9) |
| Test | No. of Slices | 71,808 | 5,840 | 2,991 |
|  | No. of Patients | 7,794 | 1,280 | 518 |
|  | Sex | 2,484M/1,794F/17U | 676M/449F/9U | 167M/93F/4U |
|  | Age | 56.0 (27.0) | 56.0 (25.0) | 60.0 (27.3) |
|  | Weight | 78.0 (23.3) | 78.0 (24.6) | 80.0 (22.0) |
|  | Height | 174.0 (16.0) | 174.0 (16.9) | 175.0 (15.0) |

*Note.*—F = female, LGE = late gadolinium enhancement, M = male, No. = number, U = unspecified. Data for age, weight, and height are presented as medians, with interquartile ranges in parentheses. Demographic data are available for a subset of patients due to incomplete metadata recording in earlier phases of the registry; therefore, demographic counts do not sum to the total number of patients in each column.

testing splits shared no institutions. Table 1 details the acquisition parameters, while Table 2 summarizes the patient demographics and dataset splits for each LGE variant.

Preprocessing involved noise pre-whitening and coil sensitivity estimation, followed by the reconstruction of reference standard MOCO PSIR images leveraging all available signal averages. These processes utilized the remote signal processing framework Tyger (12) and the open-source reconstruction framework Gadgetron (13), adhering to the PSIR LGE implementation with non-rigid respiratory motion correction, averaging, and surface coil correction (4).

## Model architecture and training

PSIRNet (Figure 1) takes as input initial coil sensitivity maps alongside noise-prewhitened, undersampled IR and PD *k*-space data from single-shot acquisitions. Since multiple IR and PD *k*-spaces were acquired over consecutive heartbeats, the first measurement pair is selected as the model input. First, the undersampled *k*-space data are transformed into the image domain via the inverse Fourier transform, yielding aliased multicoil images. Concurrently, sensitivity maps are refined using a dedicated four-level sensitivity map refinement (SMR) U-Net (17) featuring 32 first-level channels that double at each subsequent level, instance normalization (18), and LeakyReLU activations (19). The coil-combined images for IR and PD are then computed and used as the initial values in the Landweber iterations (14). Through $N = 12$ cascaded Landweber iterations, the network implicitly reconstructs IR and PD images, analogous to deep learning–based unrolled reconstruction approaches (15, 16). Each iteration applies a data-consistency step with a learned step size, followed by a joint IR–PD refinement through a four-level U-Net (17) with 96 first-level channels, which shares the same architectural design as the SMR U-Net. After the final iteration, the background phase of the PD image is removed from the IR image to preserve the magnetization sign, and surface coil correction is applied to produce the final PSIR output. The network parameters are learned end-to-end by maximizing the mean SSIM (20) between the network output and the reference standard MOCO PSIR image, eliminating the need for explicit IR–PD co-registration.

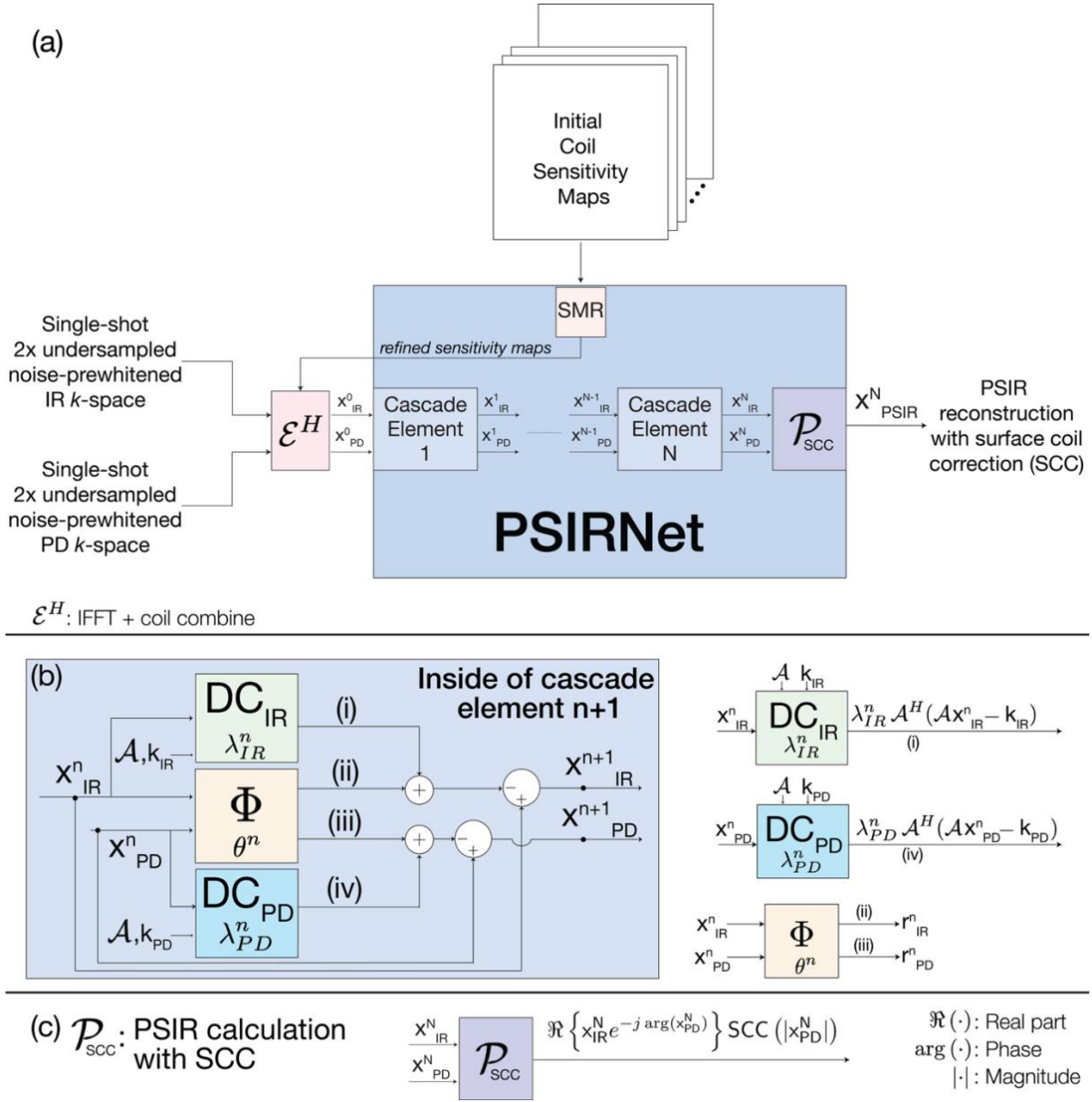

**Figure 1.** PSIRNet reconstruction. (a) Overview of PSIRNet consisting of sensitivity map refinement (SMR), $N$ cascade elements connected in series, and the final $P_{SCC}$ block calculating the PSIR image with surface coil correction (SCC). (b) Components of each element in the cascade—IR data consistency (DC), PD data consistency (DC), and joint refinement ($\Phi$)—are shown along with their inputs, their outputs, and the final aggregation of outputs. (c) Inputs and output of the $P_{SCC}$ block.

**Table 3.** PSIRNet Equations: Landweber Iterations, PSIR with SCC, and End-to-End Optimization of Network Parameters.

| Component | Equation |
|---|---|
| Landweber iterations | $\mathbf{r}_{IR}^n, \mathbf{r}_{PD}^n = \Phi(\mathbf{x}_{IR}^n, \mathbf{x}_{PD}^n; \theta^n)$ <br><br> $\mathbf{x}_{IR}^{n+1} = \mathbf{x}_{IR}^n - \lambda_{IR}^n \mathcal{A}^H(\mathcal{A}\mathbf{x}_{IR}^n - \mathbf{k}_{IR}) - \mathbf{r}_{IR}^n$ <br><br> $\mathbf{x}_{PD}^{n+1} = \mathbf{x}_{PD}^n - \lambda_{PD}^n \mathcal{A}^H(\mathcal{A}\mathbf{x}_{PD}^n - \mathbf{k}_{PD}) - \mathbf{r}_{PD}^n$ <br><br> where $\Omega \triangleq \bigcup_{n=1}^{N} \{\theta^n, \lambda_{IR}^n, \lambda_{PD}^n\}$ |
| PSIR with SCC | $\mathbf{x}_{PSIR}^N(\Omega) = \mathcal{P}_{SCC}(\mathbf{x}_{IR}^N, \mathbf{x}_{PD}^N) = \Re\left\{\mathbf{x}_{IR}^N e^{-j\,arg(\mathbf{x}_{PD}^N)}\right\} SCC(|\mathbf{x}_{PD}^N|)$ |
| End-to-end optimization | $\widehat{\Omega} = \arg\max_{\Omega} \text{SSIM}(\mathbf{x}_{PSIR}^N(\Omega), \mathbf{x}_{MOCO\ PSIR})$ |

*Note.*—Bold lowercase letters denote vectors. $\mathcal{A}$ = encoding operator, $\mathcal{A}^H$ = adjoint of the encoding operator, $\Phi$ = joint IR–PD refinement U-Net with learned parameters θ, λ = learned data-consistency step size, Ω = the set of trainable network parameters, $\mathcal{P}_{SCC}$ = PSIR with surface coil correction, $\Re$ = real part, *n* = iteration index *(n = 1, …, N)*, *N* = total number of Landweber iterations, $\mathbf{x}_{MOCO\ PSIR}$ = reference standard MOCO PSIR image, IR = inversion recovery, PD = proton density, PSIR = phase-sensitive inversion recovery, MOCO = motion corrected, SCC = surface coil correction, SSIM = structural similarity index measure.

Table 3 summarizes the Landweber iterations, the PSIR computation with surface coil correction, and the end-to-end training objective.

The full network comprises 845 million trainable parameters. The model was implemented in PyTorch version 2.8 (21) and trained using distributed data parallelism across 64 MI300X graphics processing units (GPUs) (AMD, Santa Clara, CA) on 8 nodes. We used the AdamW optimizer (22) with a learning rate of $6 \times 10^{-4}$, per-epoch decay factor of 0.98, three warmup epochs, and a weight decay of $1 \times 10^{-3}$ to learn the network parameters. The batch size was 8 per GPU, and training ran for 60,000 gradient steps. Network weights were saved every 1,250 steps, and the checkpoint with the highest mean validation SSIM was selected for inference.

### Evaluation protocol

We evaluated PSIRNet reconstructions both quantitatively and qualitatively. For quantitative evaluation, we compared PSIRNet single-shot reconstructions with reference standard MOCO PSIR images across the entire external test set using SSIM, PSNR, and NRMSE. For qualitative evaluation, two expert cardiologists (RHT and TAT, each with >10 years of experience, level III EACVI) independently scored image quality on a 5-point Likert scale ranging from 1 (non-diagnostic) to 5 (excellent confidence in LGE assessment and myocardial boundary definition) for 279 consecutive studies from the external test set. The start of this study set was randomly picked. Because a single study could include more than one LGE variant, this yielded 123 studies with bright blood, 95 with dark blood, and 84 with wideband LGE. Each multi-slice stack within a study was scored independently, resulting in 346 bright blood, 101 dark blood, and 134 wideband scores per reader. Likert scores served as the primary outcome, and quantitative metrics as secondary outcomes. Scores were analyzed separately for each reader, and when their statistical conclusions differed, the more conservative result was reported.

### Inference requirements

We quantified the inference time per slice and the peak video random access memory (VRAM) consumption for the PSIRNet model. These benchmarks were evaluated on a single A100 GPU (NVIDIA, Santa Clara, CA) and compared with the typical processing time required to generate the reference standard MOCO PSIR reconstructions using a conventional CPU-based implementation.

### Statistical analysis

Quantitative metrics are reported as mean ± standard error over the external test set, computed via macro averaging at the patient level ($n = 8,143$) to ensure equal weighting per patient. For the Likert score analysis, we first aggregated scores to the patient level by taking the mean score per patient for each reader and variant, yielding independent paired observations. We calculated mean scores across patients alongside 95% confidence intervals using the percentile bootstrap method (23) with 100,000 iterations. This approach was selected because the discrete, bounded nature of ordinal Likert scores, combined with the sample sizes per variant ($n$ = 84–123), precludes reliance on standard normal approximations for interval estimation.

To evaluate clinical performance, we formed paired differences between PSIRNet and MOCO PSIR scores. For each LGE variant, superiority of PSIRNet over MOCO PSIR was first assessed using an exact one-sided Wilcoxon signed-rank test (24). Only when superiority was not established was practical equivalence assessed using the two one-sided tests (TOST) procedure with an equivalence margin of 0.25 Likert points, corresponding to one half of the smallest measurable difference on the scoring scale, implemented via exact Wilcoxon signed-rank tests on shifted differences (25). This fixed-sequence strategy controls the per-variant Type I error rate without additional adjustment. Zero differences were handled using Pratt's method in all tests (26). Statistical significance was set at $\alpha = 0.05$. To control the family-wise error rate across the three LGE variants, Holm's step-down procedure (27) was applied at $\alpha = 0.05$. Each reader independently evaluated all image pairs, and the more conservative of the two statistical conclusions was reported, providing a conservative approach to error control across readers. All analyses were performed using R version 4.5.2 (R Foundation for Statistical Computing, Vienna, Austria) for exact Wilcoxon signed-rank tests (28, 29).

## Results

### Study population

The external test set comprised 8,143 consecutive patients. Because the raw data repository was compiled over an eight-year period (2016–2024) and demographic metadata were not routinely recorded in the earlier phases of the registry, demographic characteristics are available for 4,504 of 8,143 patients (55.3%) in the external test cohort. Among patients with available demographic data, the cohort included

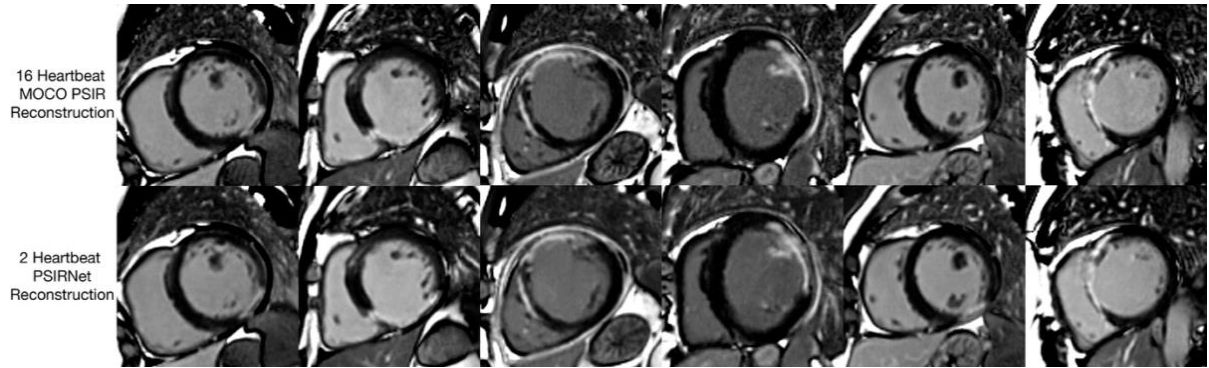

**Figure 2.** Short-axis bright blood reconstructions. At 8× acceleration, PSIRNet maintains image quality on par with the reference standard MOCO PSIR and preserves the conspicuity of the hyperenhanced myocardial infarction.

2,615 of 4,504 (58.1%) men, 1,868 of 4,504 (41.5%) women, and 21 of 4,504 (0.5%) patients with unspecified sex. The median patient age was 56.0 years (interquartile range [IQR], 27.0 years). The median patient weight was 78.0 kg (IQR, 23.3 kg), and the median patient height was 174.0 cm (IQR, 16.0 cm). From this full test set, a subset of 279 consecutive studies was evaluated in the qualitative cardiologist reader study. Within this subset, demographic characteristics were available for 203 of 279 (72.8%) patients. This representative sub-cohort comprised 129 of 203 (63.5%) men and 74 of 203 (36.5%) women, with a median age of 57.0 years (IQR, 29.5 years), median weight of 78.0 kg (IQR, 24.2 kg), and median height of 174.0 cm (IQR, 17.3 cm).

## Image quality and quantitative evaluation

PSIRNet single-shot reconstructions achieved a mean SSIM of 92.26% ± 0.04%, PSNR of 33.98 ± 0.03 dB, and NRMSE of 0.0474 ± 0.0002 relative to the reference standard MOCO PSIR images over the external test set. For bright blood LGE at 8× acceleration, PSIRNet maintained image quality on par with the MOCO PSIR reference and preserved the conspicuity of the myocardial infarction (Figure 2). In dark blood LGE at 16× acceleration, PSIRNet robustly depicted myocardial scar and borders; it overcame the severe noise present in the 2-heartbeat PSIR to achieve image quality comparable to the 32-heartbeat

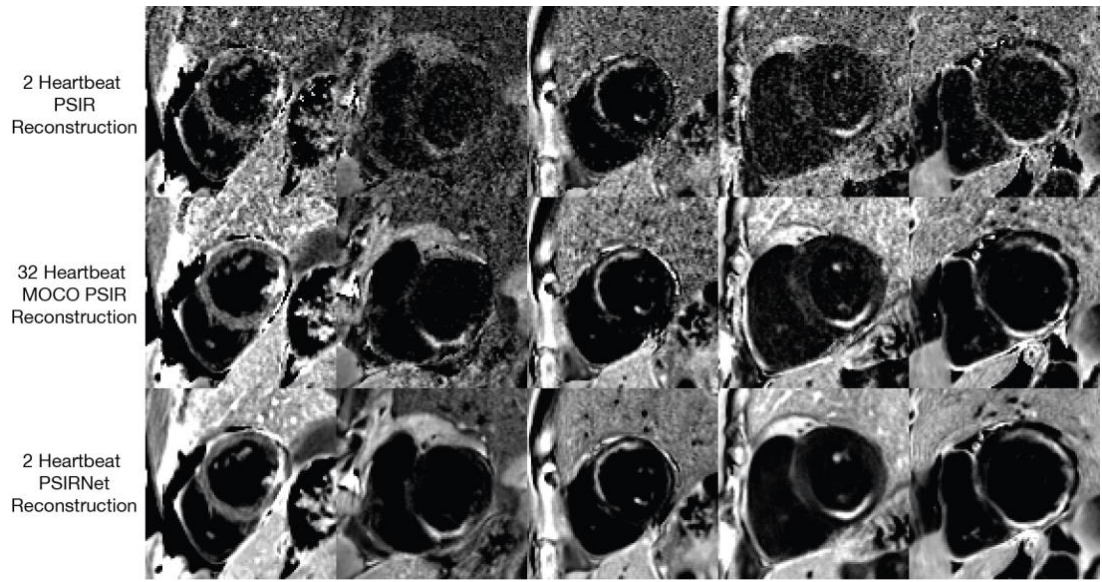

**Figure 3.** Short-axis dark blood reconstructions. PSIRNet delivers 16× acceleration with robust depiction of hyperenhanced myocardial scar and borders against the suppressed blood pool, achieving image quality comparable to or better than the reference standard MOCO PSIR.

MOCO PSIR reference (Figure 3). Similarly, while conventional wideband LGE requires extended acquisitions up to 48 heartbeats to compensate for inherently low SNR, PSIRNet achieved image quality comparable to the 24-heartbeat reference from only a 2-heartbeat acquisition (12× acceleration), demonstrating a substantial SNR improvement over the 2-heartbeat PSIR across long-axis, short-axis, and four-chamber views (Figure 4).

### Cardiologist evaluation

For dark blood LGE, PSIRNet achieved higher mean image quality scores than the MOCO PSIR reference (Reader 1: 4.46 vs 4.32; Reader 2: 4.08 vs 3.69). Bright blood scores were 4.71 vs 4.67 for Reader 1 and 4.56 vs 4.13 for Reader 2, while wideband scores were 4.44 vs 4.52 for Reader 1 and 4.01 vs 3.86 for Reader 2 (Table 4). Clinically, these scores indicate that both methods consistently provided

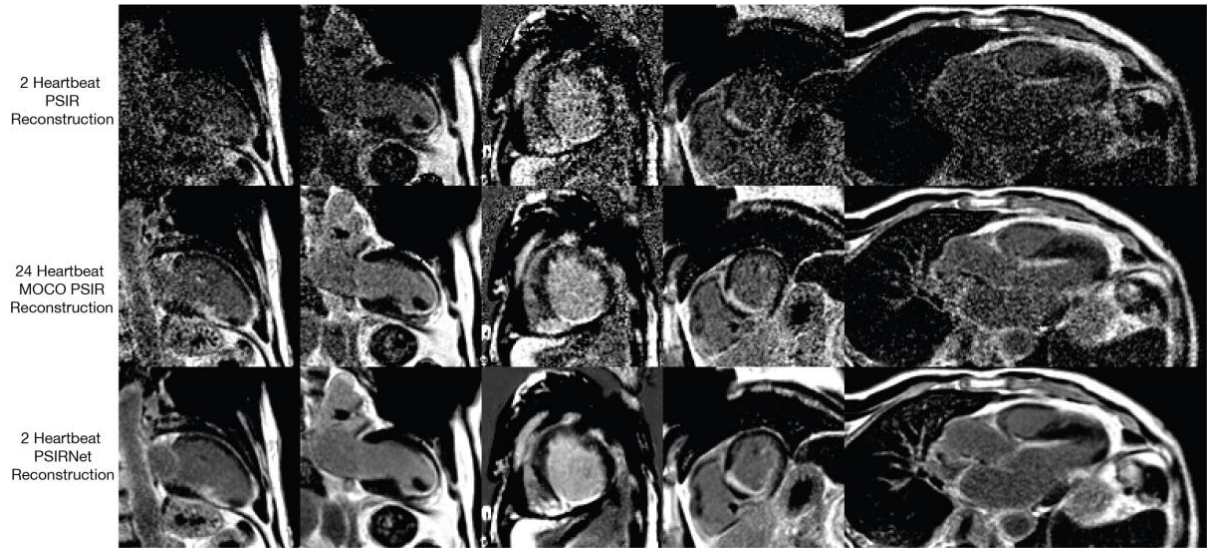

**Figure 4.** Wideband LGE reconstructions in vertical long-axis, short-axis, and four-chamber views. Wideband LGE is used in patients with implanted cardiac devices, where extended acquisitions requiring up to 48 heartbeats are necessary to compensate for inherently low SNR. PSIRNet achieves image quality comparable to or better than the 24-heartbeat MOCO PSIR reference from a 2-heartbeat acquisition (12× acceleration).

"good" to "excellent" confidence in LGE assessment and myocardial boundary definition across all variants.

In testing for superiority, PSIRNet demonstrated significantly higher image quality scores than the reference standard for dark blood LGE (conservative $P = .002$). For bright blood and wideband LGE, statistical conclusions for superiority differed between the two readers; while one reader scored PSIRNet as statistically superior (both $P < .001$), the other did not establish superiority ($P = .077$ for bright blood and $P = .813$ for wideband). Following the fixed-sequence evaluation protocol, practical equivalence was assessed for these variants using the TOST procedure with a 0.25-point margin, and the more conservative conclusion was reported. Consequently, PSIRNet was found to be statistically equivalent to MOCO PSIR for both bright blood and wideband LGE (both conservative TOST $P < .001$). All findings of superiority and equivalence remained statistically significant after applying Holm's step-down

**Table 4.** Mean Likert Scores and 95% Confidence Intervals for Image Quality of PSIRNet Reconstructions Versus Reference Standard MOCO PSIR by Reader and LGE Variant.

| Reader and Method | Bright Blood LGE | Dark Blood LGE | Wideband LGE |
|---|---|---|---|
| *Reader 1* | | | |
| PSIRNet | 4.71 (4.61, 4.79) | 4.46 (4.25, 4.64) | 4.44 (4.25, 4.61) |
| MOCO PSIR | 4.67 (4.57, 4.76) | 4.32 (4.12, 4.51) | 4.52 (4.36, 4.67) |
| *Reader 2* | | | |
| PSIRNet | 4.56 (4.48, 4.64) | 4.08 (4.01, 4.16) | 4.01 (3.90, 4.10) |
| MOCO PSIR | 4.13 (4.06, 4.20) | 3.69 (3.62, 3.77) | 3.86 (3.77, 3.94) |

*Note.*—Scores represent patient-level means across scored stacks within each study. LGE = late gadolinium enhancement, MOCO = motion corrected, PSIR = phase-sensitive inversion recovery.

procedure to control the family-wise error rate across the three LGE variants. The joint distributions of PSIRNet and MOCO PSIR scores (Figure 5) show that most paired ratings fall on or above the diagonal, consistent with the statistical findings above. Overall, the cardiologist evaluations confirmed that PSIRNet reconstructions maintained diagnostic image quality that was either practically equivalent to, or significantly better than, the reference standard across all LGE variants.

### Inference requirements

PSIRNet inference took approximately 100 msec per slice and 10 GB of VRAM on an A100 GPU. This represents a substantial acceleration compared with the reference standard MOCO PSIR, which requires more than 5 sec per slice using a conventional CPU-based implementation.

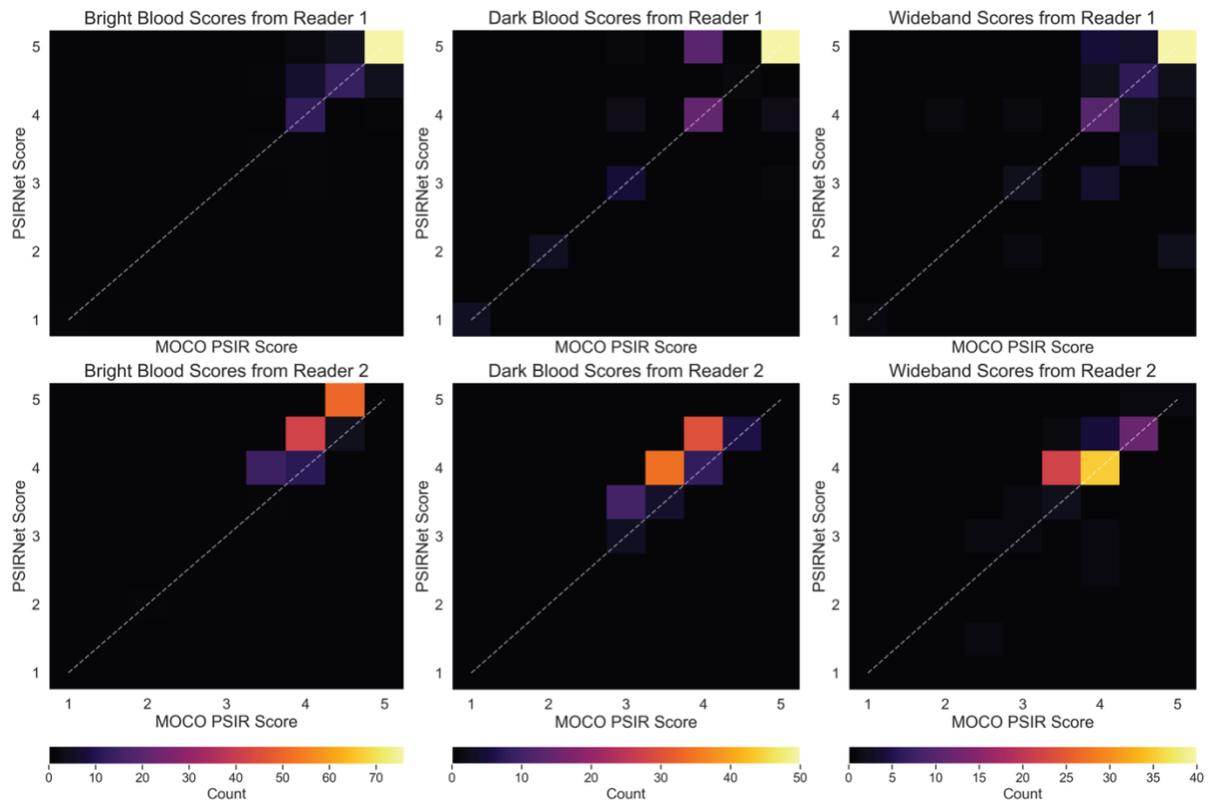

**Figure 5.** Joint distributions of PSIRNet versus MOCO PSIR Likert scores representing patient-level means across scored stacks for bright blood (left), dark blood (center), and wideband (right) LGE, shown for Reader 1 (top) and Reader 2 (bottom). Each cell represents a unique combination of PSIRNet and MOCO PSIR scores; color intensity indicates the count of paired ratings at that combination. Cells above the diagonal indicate cases where PSIRNet was scored higher; cells on the diagonal indicate agreement between methods.

# Discussion

## Main findings

Free-breathing PSIR LGE imaging is widely used to evaluate myocardial viability without the need for patient breath-holding. However, conventional protocols require extensive MOCO signal averaging to achieve a diagnostic SNR, severely limiting scan efficiency. In this study, we developed and evaluated

PSIRNet, an end-to-end, physics-guided deep learning reconstruction method. Our primary findings demonstrate that PSIRNet successfully produces diagnostic-quality PSIR LGE images from single-average, free-breathing IR and PD acquisitions, enabling an 8- to 24-fold acceleration in data acquisition. In an independent qualitative assessment, expert cardiologists found PSIRNet reconstructions to be either statistically equivalent or superior to the reference standard MOCO PSIR across bright blood, dark blood, and wideband LGE variants.

These results represent a notable progression over existing deep learning applications for LGE. Prior deep learning reconstruction methods have largely been confined to breath-held, segmented acquisitions and tested on modest, single-center cohorts, which limits their generalizability. By leveraging a large-scale, multi-institutional dataset of over 800,000 slices encompassing varied field strengths, scanner models, software versions, and LGE variants, PSIRNet demonstrates robust generalization.

### Clinical and workflow implications

The transition from an extensive, multi-average acquisition to a single-shot deep learning reconstruction carries substantial implications for cardiac MRI workflows. By achieving statistically equivalent or superior diagnostic quality from a 2-heartbeat acquisition, PSIRNet markedly reduces the time patients must spend inside the scanner. This acceleration can be leveraged in multiple ways: it allows for thinner slices to improve spatial resolution, enables comprehensive whole-heart volumetric coverage within a practical timeframe, frees up scanner time to accommodate higher patient throughput, or permits the inclusion of otherwise optional diagnostic sequences in the imaging protocol.

Additionally, PSIRNet streamlines the computational pipeline. Relying on a conventional CPU-based implementation, non-rigid motion correction, registration, and averaging can take more than 5 sec of processing time per slice, leading to noticeable delays at the scanner console before images are available for review. PSIRNet mitigates this computational bottleneck, reducing inference time to approximately 100 msec per slice by leveraging GPU acceleration. This rapid reconstruction allows technologists to verify image quality and diagnostic adequacy in real-time, minimizing the need for patient callbacks or repeat sequences.

## Limitations

This study has several limitations. First, only retrospective performance was evaluated, and the network has not yet been deployed in a routine clinical workflow. Prospective validation is required to confirm that the performance observed here translates to true single-shot acquisitions and to allow real-time evaluation by clinicians making diagnostic decisions. Second, although the dataset spans multiple institutions, scanner models, software versions, and field strengths, all data were acquired on Siemens scanners (Siemens Healthineers, Erlangen, Germany). Generalizability to other scanner vendors remains to be established. Third, the MOCO PSIR reconstruction served as both the training target and the evaluation reference, inherently limiting the network's ability to surpass the quality of its own reference and propagating any systematic errors present in the conventional pipeline. Finally, our evaluation focused on image quality metrics and expert Likert scores but did not assess downstream clinical tasks such as scar quantification or treatment decision-making; equivalent perceptual quality does not guarantee equivalent clinical utility.

## Conclusion

PSIRNet produces diagnostic-quality free-breathing PSIR LGE images from a single interleaved IR/PD acquisition over two heartbeats, achieving image quality that is statistically equivalent or superior to conventional MOCO PSIR reconstructions across bright blood, dark blood, and wideband LGE variants. With two-heartbeat imaging and 100-msec inference time, PSIRNet enables more slices, greater volumetric heart coverage, and much faster clinical workflows. Prospective validation and evaluation of downstream clinical tasks are needed to establish clinical readiness.

## Code and Model Availability

The source code and trained model weights are publicly available at [github.com/microsoft/psirnet](github.com/microsoft/psirnet) and [huggingface.co/microsoft/psirnet](huggingface.co/microsoft/psirnet), respectively.